\RequirePackage{fix-cm}
\documentclass[smallextended,natbib]{svjour3}       
\smartqed  
\usepackage[misc,geometry]{ifsym} 
\usepackage{graphicx,amsmath}
\usepackage[dvipsnames]{xcolor}
\usepackage{appendix}
\begin{document}

\title{Laplacian Matrices for Extremely Balanced and Unbalanced Phylogenetic Trees
}
%



\author{T. Ara\'ujo Lima         \and
        Marcus A. M. de Aguiar 
}


\institute{
 \begin{acknowledgements}
  This work was carried out with the support of National Council for Scientific and Technological Development -- CNPq, Brazil (Cases No. 158002/2018-0 and 302049/2015-0).
 \end{acknowledgements}
              T. Ara\'ujo Lima (\Letter) \at
              \email{tiagoapl@ifi.unicamp.br}           
           \and
           Marcus A. M. de Aguiar \at
              \email{aguiar@ifi.unicamp.br}\\
              \newline
              Instituto de F\'{i}sica ‘Gleb Wataghin’, Universidade Estadual de Campinas, Unicamp, 13083-970 Campinas, SP, Brazil
}

\date{Received: date / Accepted: date}

\maketitle

\begin{abstract}

Phylogenetic trees are important tools in the study of evolutionary relationships between species. Measures such as the index of Sackin, Colless, and Total Cophenetic have been extensively used to quantify tree balance, one key property of phylogenies. Recently a new proposal has been introduced, based on the spectrum of the Laplacian matrix associated with the tree. In this work, we calculate the Laplacian matrix analytically for two extreme cases, corresponding to fully balanced and fully unbalanced trees. For maximally balanced trees no closed form for the Laplacian matrix was derived, but  we present an algorithm to construct it. We show that Laplacian matrices of fully balanced trees display self-similar patterns that result in highly degenerated  eigenvalues. Degeneracy is the main signature of this topology, since it is totally absent in fully unbalanced trees. We also establish some analytical and numerical results about the largest eigenvalue of Laplacian matrices for these topologies.

\end{abstract}

\section{Introduction}
\label{intro}

\ \ \ \ \ Phylogenetic trees are key tools in evolutionary theory, allowing species to be organized by genetic or phenotypic similarity \citep{fel:2004}. The tips of the tree correspond to living (extant) species, internal nodes indicate speciation events and branch lengths the duration of species between their extinction (interrupted branch) or speciation (bifurcation into two or more branches). Besides these direct information, several analytical and computational methods have been developed to understand and characterize other aspects of phylogenies, related to their topology and distribution of branch lengths. These studies are motivated by the belief that structural properties of a tree reflect, to some extent, the processes of speciation underlying its formation \citep{nee:1992,pen:2013,mor:2014,cab:2017,cos:2019}. One important piece of geometric information has to do with the tree's balance, measured by indices such as the Sackin \citep{sac:1972,sha:1990,fis:2019}, Colless \citep{col:1982,cor:2020}, and Total Cophenetic \citep{mir:2013,car:2013}.\\

Recent works have suggested that the Laplacian matrix associated with the phylogenetic tree can provide useful information about its structural properties in a concise form \citep{lew:2016,lew:2016b}. These authors consider a modified version of the Laplacian matrix (see next section) and compute the smoothed  spectral density function to generate a density profile of the spectrum. They argue that the density profile kurtosis contains information about the tree balance. A flat peak indicates that the tree is balanced, whereas a steep peak means the tree is unbalanced. Motivated by this proposal, we here calculate the Laplacian matrix analytically for extreme types of trees, corresponding to fully and maximally balanced trees and also fully unbalanced trees of arbitrary sizes. We study the spectrum of these matrices, looking for signatures of balance properties in eigenvalues of the Laplacian, as suggested by \citep{lew:2020}. Analytic knowledge of these cases might help understand and classify more typical trees found in nature.\\

Here we consider only rooted binary trees where tips represent extant species and internal nodes indicate most recent common ancestors, serving as proxies for speciation events. It is convenient to call the internal nodes {\it ancient species} that eventually branched into the living ones. Since we are mostly interested in the shape of the trees we shall fix the distance between consecutive speciation events to 1. If $n$ is the number of tips, then the total number of species in the tree, extant plus ancient, is $N=2n-1$. Traditionally, the Laplacian matrix $\boldsymbol{\Lambda}$ for an un-directed network with $N$ nodes is defined as $\boldsymbol{\Lambda} = \boldsymbol{K} - \boldsymbol{A}$ where the Adjacency matrix $\boldsymbol{A}$ has elements
\begin{equation}
	\label{a}
	a_{i,j}=\left\{ \begin{array}{l}
		1, \quad \textrm{if nodes $i$ and $j$ are connected}\\
		0, \quad \textrm{if nodes $i$ and $j$ are not connected,}\end{array}\right. 
\end{equation}
and the Degree matrix $\boldsymbol{K}$ is a diagonal matrix with entries
\begin{equation}
	\label{k}
	k_i=\sum_{j=1}^N a_{i,j},
\end{equation}
corresponding to the degrees of the nodes. Both of these matrices are symmetric.\\

In the next section we define two modified Laplacian matrices considered in \citep{lew:2016} and introduce smaller, reduced matrices, based on extant species only. In section 3 we show how these matrices and their eigenvalues can be computed analytically in the cases of maximal and minimal balance. We find that fully balanced trees display self-similar patterns that result in highly degenerated eigenvalues, whereas fully unbalanced trees have no degeneracy and each eigenvalue is unique. Using a set of randomly generated trees we present numerical evidence that the eigenvalues of fully unbalanced and maximally balanced trees are the largest and smallest possible eigenvalues respectively, setting limits on the spectrum of typical trees. In section 4 we summarize our conclusions.

\begin{figure*}[!htpb]
 \centering
 \includegraphics[width=0.9\textwidth]{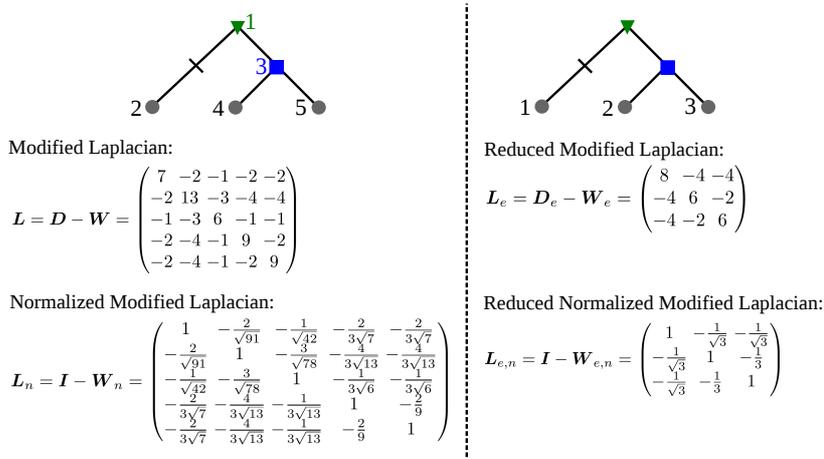}
 \caption{Examples of modified versions of Laplacian matrices discussed in this paper. Green triangles represent the root. The blue squares are ancient species, and gray circles are extant species. The black tick indicates that branch length is increased by 1.}
 \label{scheme}
\end{figure*}

\section{Modified Laplacian Matrices}

\ \ \ \ \ In this session we describe the two modified versions of the Laplacian matrix presented in \citep{lew:2016} and introduce a {\it reduced} form that takes into account only the extant species. Figure \ref{scheme} shows an illustration with all definitions and a straightforward example.\\

In the first version, we replace the Adjacency matrix $\boldsymbol{A}$ by the Distance matrix $\boldsymbol{W}$ with elements $w_{i,j}$ given by the total branch length between nodes $i$ and $j$. These lengths represent the temporal distance between subsequent speciation events or between an extant species and the speciation event that gave rise to it. Accordingly, the Degree matrix is replaced by another diagonal matrix, $\boldsymbol{D}$, with entries
\begin{equation}
	\label{degree}
	d_i=\sum_{j=1}^N w_{i,j}.
\end{equation}
The Modified Laplacian matrix is finally defined by $\boldsymbol{L}=\boldsymbol{D}-\boldsymbol{W}$.\\

The second modified version is constructed as follows: we define the matrix $\boldsymbol{D}^{-1/2}$ such that $\boldsymbol{I}=\boldsymbol{D}^{-1/2}\boldsymbol{D}\boldsymbol{D}^{-1/2}$ where  $\boldsymbol{I}$ is the identity matrix. Then, the Normalized Modified Laplacian matrix is $\boldsymbol{L}_n = \boldsymbol{D}^{-1/2} \boldsymbol{L} \boldsymbol{D}^{-1/2}  = \boldsymbol{I} - \boldsymbol{W}_n$.  Its matrix elements $\ell_{i,j}$ are given by
\begin{equation}
	\label{nL}
	\ell_{i,j} = \left\{ \begin{array}{l}
		1, \quad \textrm{if $i=j$}\\
		-\frac{w_{i,j}}{\sqrt{d_i d_j}}, \quad \textrm{if $i\neq j$.}\end{array}\right.
\end{equation}

In the next sections, we will construct these two modified versions of the Laplacian matrix for the fully, maximally balanced, and fully unbalanced trees with an arbitrary number of species $N$. These cases are of fundamental importance for the balance analysis of phylogenetic trees \citep{lew:2016,lew:2016b}. We will show that Distance matrices for fully balanced trees exhibit self-similar patterns that make their calculation possible.\\

Finally, we construct reduced Laplacian matrices $\boldsymbol{L}_e = \boldsymbol{D}_e-\boldsymbol{W}_e$ and its normalized version $\boldsymbol{L}_{e,n} = \boldsymbol{I} - \boldsymbol{W}_{e,n}$, characterized only by the tree tips  \citep{lew:2020}. In this representation the matrices are $n \times n$, nearly the half size of the full matrix, as illustrated in Figure \ref{scheme}. The reduced Distance matrix $\boldsymbol{W}_e$ is identical to $\boldsymbol{W}$, but restricted to the tips. The Degree matrix then becomes
\begin{equation}
 \label{degreer}
 d_{e,i}=\sum_{j=1}^n w_{e,i,j}.
\end{equation}
Normalization is defined similarly to Eq.(\ref{nL}), with $w_{i,j}$ replaced by $w_{e,i,j}$ and $d_i$ by $d_{e,i}$. The self-similar patterns found in the distance matrices $\boldsymbol{W}$ and $\boldsymbol{W}_n$ persists in $\boldsymbol{W}_e$ and $\boldsymbol{W}_{e,n}$ for fully balanced trees, also resulting in high level of degeneracy in the matrix's spectrum and being a signature of tree balance. 

\section{Results}

\subsection{Fully Unbalanced Trees}
\label{lapmat}

\subsubsection{Matrices}
\label{asym}

\begin{figure*}[!htpb]
 \centering
 \includegraphics[width=0.9\columnwidth]{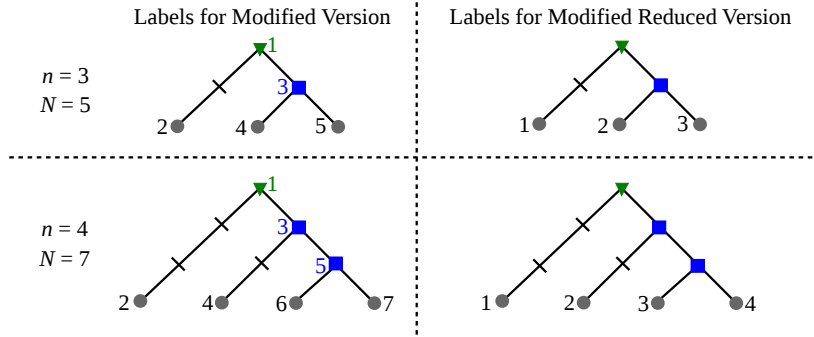}
 \caption{Fully unbalanced trees for $n=3$ and 4. Green triangles represent the root. The blue squares are ancient species, and gray circles are extant species. The black tick indicates that branch length is increased by 1. Left Panels: Labels for the Modified Laplacian Matrices. Odd numbers label ancient species, except for node $N$, which is extant. Right Panels: Labels for the modified reduced version, where only extant species are labeled nodes.}
 \label{trees_unbal}
\end{figure*}

\begin{figure*}[!htpb]
 \centering
 \includegraphics[width=0.9\columnwidth]{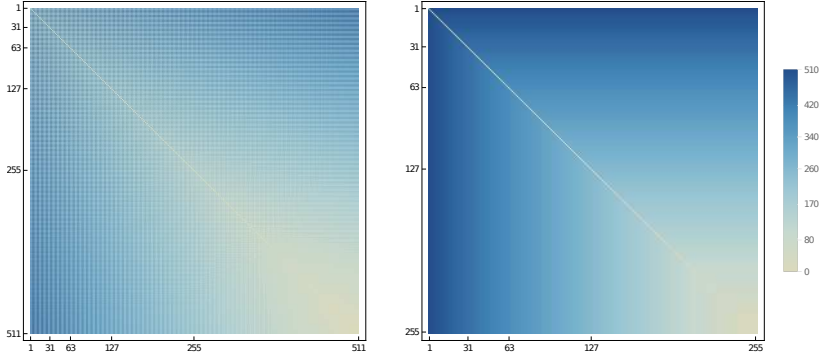}
 \caption{$\boldsymbol{W}^{(256)}$, distance matrix for a fully unbalanced tree with $n=256$ leaves and $N=511$ nodes. The small distances are abundant for large $(i,j)$. Those nodes represent the most recent speciation events, so they are more close to each other. Left Panel: Modified matrix version. Right Panel: Reduced modified version.}
 \label{mat_plot2}
\end{figure*}

Fully unbalanced trees (also called caterpillars) are characterized by having all speciation events resulting in one extant species and one branch that speciates again, except for the last event that gives rise to two extant species, as illustrated in Figure \ref{trees_unbal}. We will use the number of extant species $n=2,3,4,...$ to label the matrices and their elements. We start by the Modified Laplacian matrices version. The reduced version will be treated in sequence, where we present the eigenvalues for these matrices. For the trees in Figure \ref{trees_unbal}, $\boldsymbol{L}^{(n)}$ are given by:
\begin{equation}
 \label{Ls_unb}
 \boldsymbol{L}^{(3)} =\begin{pmatrix}
  7 & -2 & -1 & -2 & -2\\
  -2 & 13 & -3 & -4 & -4\\
  -1 & -3 & 6 & -1 & -1\\
  -2 & -4 & -1 & 9 & -2\\
  -2 & -4 & -1 & -2 & 9\end{pmatrix}, \quad
 \boldsymbol{L}^{(4)} =\begin{pmatrix}
  15 & -3 & -1 & -3 & -2 & -3 & -3\\
  -3 & 30 & -4 & -6 & -5 & -6 & -6\\
  -1 & -4 & 12 & -2 & -1 & -2 & -2\\
  -3 & -6 & -2 & 22 & -3 & -4 & -4\\
  -2 & -5 & -1 & -3 & 13 & -1 & -1\\
  -3 & -6 & -2 & -4 & -1 & 18 & -2\\
  -3 & -6 & -2 & -4 & -1 & -2 & 18\end{pmatrix}.
\end{equation}
In order to derive general expressions for $\boldsymbol{L}^{(n)}$ we first consider the Distance matrix $\boldsymbol{W}^{(n)}$ with elements $w_{i,j}^{(n)}$. Because $\boldsymbol{W}^{(n)}$ is a symmetric matrix, we only need to calculate the elements with $j>i$. Tree nodes are labeled like in Figure \ref{trees_unbal}, and this is of paramount importance to obtain simple expressions for the matrix elements. The labeling splits the nodes into two kinds according to their parity:  species labeled with odd numbers, except for node $N$, correspond to ancient species while those with even numbers are extant species. By inspecting small matrices (see Appendix \ref{AA1}) we find that, for $j > i$, 
\begin{equation}
 \label{wij_unbal}
 w_{i,j}^{(n)}=\left\{ \begin{array}{l}
  \frac{j-i}{2}, \quad \text{odd } i \text{ and } j\\
  2n-i, \quad \text{even } i \text{ and } j\\
  n-\frac{i+1}{2}, \quad \text{odd } i \text{ and even } j\\
  n-i+\frac{j+1}{2}, \quad \text{even } i \text{ and odd } j.\end{array}\right.
\end{equation}
Except for odd-odd case, all expressions depend on $n$, since branch lengths increase as the tress grows. A simple change in $i$ and $j$ provides the elements below the diagonal. The left panel of Figure \ref{mat_plot2} shows a heat map plot of $\boldsymbol{W}^{(256)}$.\\

The Modified Degrees matrix $\boldsymbol{D}^{(n)}$, a diagonal matrix whose elements are obtained from Eq. (\ref{degree}) is also calculated in Appendix \ref{AA1}. Once again parity plays a key role in the calculation. We obtain
\begin{equation}
 \label{di_unbalf}
 2d_i^{(n)} = \left\{ \begin{array}{l}
  3n^2-n(2i+3)+i(i+1), \quad i \text{ odd}\\
  7n^2-n(4i+7)+i(i+2), \quad i \text{ even.}\end{array}\right.
\end{equation}
Once $w_{i,j}^{(n)}$ and $d_i^{(n)}$ have been calculated, the Normalized Modified matrix is obtained using (\ref{wij_unbal}) and (\ref{di_unbalf}) into (\ref{nL}).

Now, we consider the Reduced Modified version of Laplacian matrices $\boldsymbol{L}^{(n)}_e$. For $n=3$ and $4$ they are given by:
\begin{equation}
 \label{Ls_unb_l}
 \boldsymbol{L}^{(3)}_e =\begin{pmatrix}
  8 & -4 & -4\\
  -4 & 6 & -2\\
  -4 & -2 & 6\end{pmatrix}, \quad
 \boldsymbol{L}^{(4)}_e =\begin{pmatrix}
  18 & -6 & -6 & -6\\
  -6 & 14 & -4 & -4\\
  -6 & -4 & 12 & -2\\
  -6 & -4 & -2 & 12\end{pmatrix}.
\end{equation}
To compute the reduced matrix for an arbitrary number of species, we again start with the Distance matrix $\boldsymbol{W}^{(n)}_e$.  From (\ref{Ls_unb_l}) we see that $w_{i,j}^{(n)}$ is independent of $j$ for $j>i$, depending only on $i$ and $n$ (indicating the growth of the branch lengths when the tree increases). We find
\begin{equation}
 \label{wij_unbal_l2}
 w_{i,j}^{(n)} = \left\{ \begin{array}{l}
  2(n-i), \quad j>i\\
  2(n-j), \quad i>j.\end{array}\right.
\end{equation}
The right panel of Figure \ref{mat_plot2} shows the Distance matrix $\boldsymbol{W}^{(256)}_e$. The Modified Degree matrix $\boldsymbol{D}^{(n)}_e$ is  given by Eq.(\ref{degreer}). Splitting the sum over $j$ into $i>j$ and $i<j$ we get
\begin{equation}
 \label{di_unbal_l2}
 d_i^{(n)} = 2n^2 - (2n-i)(i+1).
\end{equation}
This completes the information needed to build the Modified Laplacian matrix for extant species and the normalized version.

\subsubsection{Spectrum}

\ \ \ \ \ The phylogenetic analysis presented in \citep{lew:2020} is based on the eigenvalues of the reduced Laplacian matrix associated with the corresponding tree. Here, we show how the spectrum of eigenvalues varies with tree size in the unbalanced case. We are interested in the set $\{\lambda_i\}_{i=1}^n$ such that
$\boldsymbol{L}_e \boldsymbol{v}_i = \lambda_i \boldsymbol{v}_i$ where $\boldsymbol{v}_i$ is the eigenvector associated to eigenvalue $\lambda_i$. In Appendix \ref{AA2} we show that $\boldsymbol{L}_e$ is a positive semi-definite matrix, i.e., $\boldsymbol{x}^t \boldsymbol{L}_e \boldsymbol{x} \geq 0$, for any vector $\boldsymbol{x}$ and its transpose $\boldsymbol{x}^t$. Consequently $\lambda_i \geq 0$. Suppose that $\lambda=0$ is an eigenvalue of $\boldsymbol{L}_e$ with eigenvector $\boldsymbol{y}$, we get from (\ref{vLvfinal}):
\begin{equation}
 \label{lamb0}
 \boldsymbol{y}^t \boldsymbol{L}_e \boldsymbol{y} = \sum_{p>q} (y_p - y_q)^2 w_{p,q} = 0
\end{equation}
The expression above only can be satisfacted if $y_p=y_q \forall (p,q)$. Then, $\boldsymbol{y}^t = (1,1,1,...,1)$ is the eigenvector of $\boldsymbol{L}_e$ for $\lambda=0$ and this eigenvalue is unique. The spectrum always contains a null eigenvalue whereas the others are real and positive:
\begin{equation}
 \label{eigenz}
 \{\lambda_i\}_{i=1}^n = \{0, \lambda_2, \lambda_3, ..., \lambda_n\},
\end{equation}
with $ 0<\lambda_2\leq\lambda_3\leq...\leq\lambda_n$.  For the fully unbalanced case we obtain (see Appendix \ref{AA3})
\begin{equation}
 \label{lambr_unbf2}
 \lambda_{r+1} = n(n-1)+r(r+1), \qquad r=1,2,3,...,n-1.
\end{equation}
Figure \ref{lambplot} shows the behavior of the largest eigenvalue $\lambda_n=2n(n-1)$ with the tree size and will be discussed later.

\subsection{Fully Balanced Trees}
\label{sym}

\subsubsection{Matrices}

\begin{figure*}[!htpb]
 \centering
 \includegraphics[width=0.9\columnwidth]{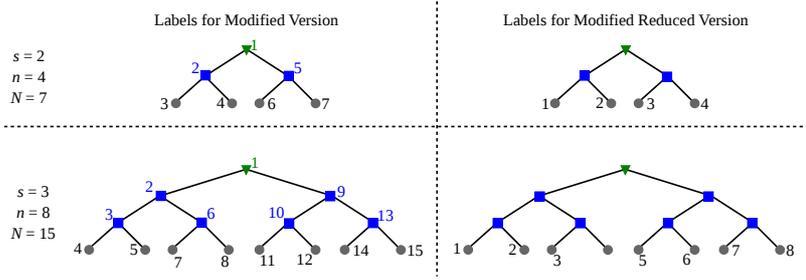}
 \caption{Fully balanced trees for $s=2(n=4)$ and $s=3(n=8)$. Green triangles represent the root. The blue squares are ancient species, and gray circles are extant species. Left Panels: Labels for the Modified Laplacian Matrices. Right Panels: Labels for the modified reduced version, where only extant species are labeled nodes.}
 \label{trees_bal}
\end{figure*}
\begin{figure*}[!htpb]
 \centering
 \includegraphics[width=0.75\columnwidth]{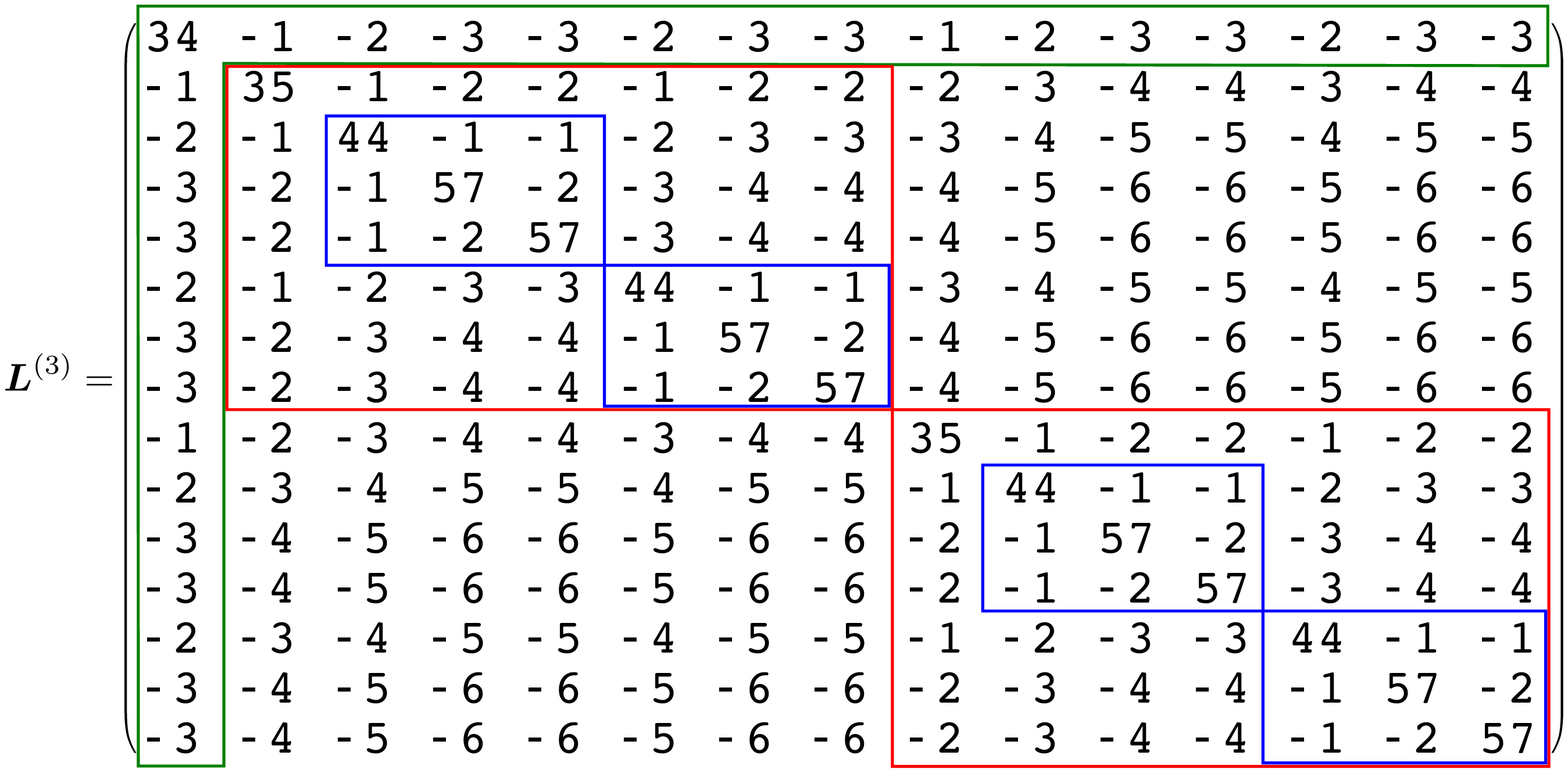}
 \caption{Laplacian matrix $\boldsymbol{L}^{(3)}$. Each step in $s$ increases the matrix regarding a self-similar pattern. Each red larger (blue smaller) sub-matrix is identical. The first row and column are highlighted in green.}
 \label{mat_scheme}
\end{figure*}
\begin{figure*}[!htpb]
 \centering
 \includegraphics[width=0.9\textwidth]{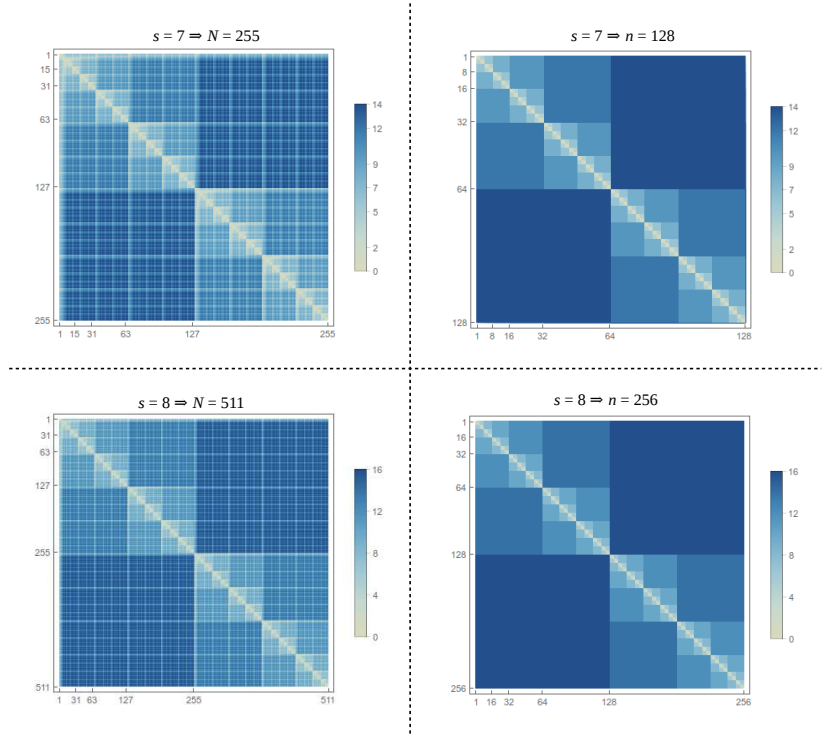}
 \caption{Self-similar pattern of the distance matrix $\boldsymbol{W}^{(s)}$ for the fully balanced trees and some depths. Dark blue cells indicate large distances while the light blue ones indicate small distances. The large distances are concentrated in the filling sub-matrices $\boldsymbol{W}^{(s)}_F$, distances between nodes on opposite sides of the root node. The small distances are concentrated in regions that come from distance matrix $\boldsymbol{W}^{(s-1)}$, distances between nodes on the same side of the root node. Left Panels: Modified matrices versions. Right Panel: Reduced modified versions.}
 \label{mat_plot}
\end{figure*}


\ \ \ \ \ A tree can be fully balanced only if number of extant species satisfies $n=2^s$, for $s=0,1,2,...$. For $n\neq 2^s$, the tree can be maximally balanced, and we discuss this case in the next section. Unlike the fully unbalanced case, multiple speciation events occur at the same time. Because $n=2^s$, we label the matrices and their elements by $s$, that represents the tree depth, the distance between the root node and the tips. Figure \ref{trees_bal} shows two examples. As in the previous section, we start by computing the Modified Laplacian matrices and later calculate their reduced versions. For the trees in Figure \ref{trees_bal}, $\boldsymbol{L}^{(2)}$ is given in Eq. (\ref{Ls_bal}) and $\boldsymbol{L}^{(3)}$ is shown in Figure \ref{mat_scheme}, where we illustrate the ingredients discussed below.
\begin{equation}
 \label{Ls_bal}
 \boldsymbol{L}^{(2)} =\begin{pmatrix}
  10 & -1 & -2 & -2 & -1 & -2 & -2\\
  -1 & 11 & -1 & -1 & -2 & -3 & -3\\
  -2 & -1 & 16 & -2 & -3 & -4 & -4\\
  -2 & -1 & -2 & 16 & -3 & -4 & -4\\
  -1 & -2 & -3 & -3 & 11 & -1 & -1\\
  -2 & -3 & -4 & -4 & -1 & 16 & -2\\
  -2 & -3 & -4 & -4 & -1 & -2 & 16\end{pmatrix}.
\end{equation}
A look at these matrices shows that $\boldsymbol{L}^{(s)}$ presents a self-similar pattern, with the Laplacian  for a given $s$ containing information of the matrices for smaller $s$. In Figure \ref{mat_scheme}, the sub-matrices in red are exactly the Distance matrix represented in Eq. (\ref{Ls_bal}) (recall that the diagonal comes from de Degrees matrix), while the sub-matrices in blue represent cherries, trees with only two extant species.  We can extend this idea including the trivial case $s=0 \Rightarrow n=1 \Rightarrow N=1 \Rightarrow \boldsymbol{L}^{(0)}=(0)$, a $1 \times 1$ matrix. To obtain a general rule for the matrix elements we start with the Distance matrix constructing a recursive rule for $\boldsymbol{W}^{(s)}$. We write
\begin{equation}
 \label{Ws}
 \boldsymbol{W}^{(s)} = \begin{pmatrix}
  0 & \ldots & w_{1,N}^{(s)}\\
  \vdots & \boldsymbol{W}^{(s-1)} & \boldsymbol{W}_F^{(s)}\\
  w_{N,1}^{(s)} & \boldsymbol{W}_F^{(s)} & \boldsymbol{W}^{(s-1)}\end{pmatrix}.
\end{equation}
To complete the matrix, we need the first line, equal to the first column, and the {\it filling} sub-matrices $\boldsymbol{W}_F^{(s)}$. Details of this construction are in Appendix \ref{AA4}, where examples are also provided. The self-similar pattern is illustrated in Figure \ref{mat_plot} for $s=7$ and 8. Large distances are concentrated in the filling sub-matrices $\boldsymbol{W}^{(s)}_F$, representing distances between nodes on opposite sides of the root. Small distances, on the other hand, are concentrated in $\boldsymbol{W}^{(s-1)}$, accounting for nodes on the same side of the root.
\\

The next step is to obtain the Modified Degree matrix $\boldsymbol{D}^{(s)}$. Since the first line of $\boldsymbol{W}^{(s)}$ is special, we start with node 1. Inspecting Eqs. (\ref{Ls_unb_A}), (\ref{Ls_bal}), (\ref{W1}) and Figure \ref{mat_scheme} we see that 
\begin{equation*}
 d_{1}^{(s)}=2 \cdot 1 + 4 \cdot 2 + 8 \cdot 3 + 16 \cdot 4 + ... = (2^1) \cdot 1 + (2^2) \cdot 2 + (2^3) \cdot 3 + (2^4) \cdot 4 + ...
\end{equation*}
\begin{equation}
 \label{d1f}
 \qquad = \sum_{m=1}^s m2^m = 2 [1 + 2^s(s-1)].
\end{equation}
The computation of the modified degrees of the remaining nodes is in Appendix \ref{AA4}, resulting in
\begin{equation}
 \label{di_bal_final2}
 d_i^{(s)} = \left\{ \begin{array}{l}
  d_{i-1}^{(s-1)} + 2^s(w_{1,i}^{(s)}+s-1) + 1, \quad 2 \leq i \leq 2^s\\
  d_{i+1-2^s}^{(s)}, \quad 2^s+1 \leq i \leq 2^{s+1}-1.\end{array}\right.
\end{equation}
As in the fully unbalanced case in section \ref{asym} the knowledge of all matrix elements $w_{i,j}^{(s)}$ and $d_i^{(s)}$ also allows us to write the Normalized Modified Laplacian matrices.\\

Now, we consider the Reduced Modified version of Laplacian matrices $\boldsymbol{L}^ {(s)}_e$. For $s=2$ and 3 they are given by:
\begin{equation}
 \label{Ls_bal_l}
 \boldsymbol{L}^{(2)}_e =\begin{pmatrix}
  10 & -2 & -4 & -4\\
  -2 & 10 & -4 & -4\\
  -4 & -4 & 10 & -2\\
  -4 & -4 & -2 & 10\end{pmatrix}, \quad
  \boldsymbol{L}^{(3)}_e =\begin{pmatrix}
  34 & -2 & -4 & -4 & -6 & -6 & -6 & -6\\
  -2 & 34 & -4 & -4 & -6 & -6 & -6 & -6\\
  -4 & -4 & 34 & -2 & -6 & -6 & -6 & -6\\
  -4 & -4 & -2 & 34 & -6 & -6 & -6 & -6\\
  -6 & -6 & -6 & -6 & 34 & -2 & -4 & -4\\
  -6 & -6 & -6 & -6 & -2 & 34 & -4 & -4\\
  -6 & -6 & -6 & -6 & -4 & -4 & 34 & -2\\
  -6 & -6 & -6 & -6 & -4 & -4 & -2 & 34\end{pmatrix}.
\end{equation}
The self-similar pattern verified in the Distance matrices $\boldsymbol{L}^{(s)}$ appears again. The structure of $\boldsymbol{W}^{(s)}_e$ is now quite simple:
\begin{equation}
 \label{Ws_l}
 \boldsymbol{W}^{(s)}_e = \begin{pmatrix}
  \boldsymbol{W}^{(s-1)}_e & \boldsymbol{W}_{e,F}^{(s)}\\
  \boldsymbol{W}_{e,F}^{(s)} & \boldsymbol{W}^{(s-1)}_e\end{pmatrix},
\end{equation}
where the elements of the filling sub-matrices are
\begin{equation}
 \label{wf_l}
 w_{i,j}^{(s)}=2s, \quad \forall w_{i,j} \in \boldsymbol{W}_{e,F}^{(s)}.
\end{equation}
The self-similar pattern of $\boldsymbol{L}^{(s)}$ is shown in Figure \ref{mat_plot}.\\

The next step is to obtain the Modified Degree matrix for extant species $\boldsymbol{D}^{(s)}_e$. We note that the modified degrees are constant, independent of the node:
\begin{equation}
 \label{di_l}
 d_i^{(s)}=d^{(s)}=\sum_{j=1}^n w_{i,j}^{(s)}.
\end{equation}
By inspection, we see that
\begin{equation*}
 d^{(s)} = 2 \cdot 1 + 4 \cdot 2 + 6 \cdot 4 + 8 \cdot 8 + ...
\end{equation*}
or
\begin{equation}
\label{d1f_l}
 d^{(s)} =\sum_{m=1}^s m2^m = 2 [1 + 2^s(s-1)].
\end{equation}
These results coincide with Eq. (\ref{d1f}), the modified degree of the root node for the complete matrix. This means that the sum of the distances from the root to all nodes (internal and leaves) is the same as the sum of all distances between tips in the fully balanced case for same $s$.

\subsubsection{Spectrum}

\ \ \ \ \ For the fully balanced case, the sets of eigenvalues of $\boldsymbol{L}_e$ are given by
\begin{equation}
 \label{lambs_balf2}
 \lambda_r = 2 [2^r+2^s(s-1)], \quad r=1,2,3,...,s,
\end{equation}
with multiplicity $2^{s-r}$. This result is shown in Appendix \ref{AA5}. Figure \ref{lambplot} shows the behavior of the largest eigenvalue $\lambda_s=s2^{s+1}$ with the tree size.

\subsection{Maximally Balanced Trees}
\label{maxbal}

\subsubsection{Matrices}

\begin{figure*}[!htpb]
 \centering
 \includegraphics[width=0.9\columnwidth]{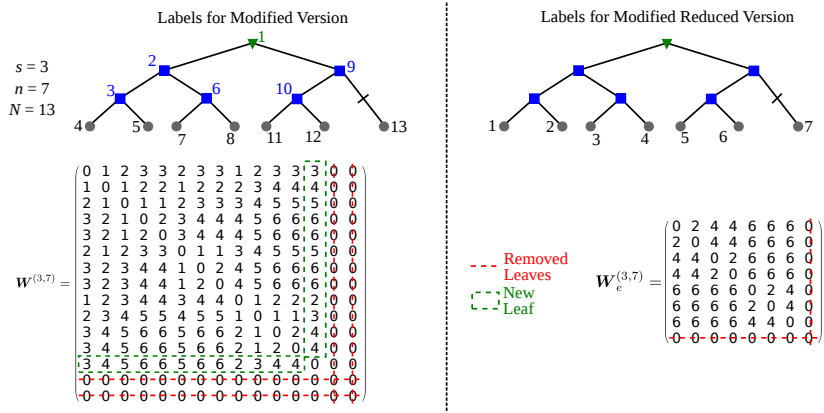}
 \caption{Maximally balanced tree with seven tips. Just one step reaches the desired tree starting the algorithm with a tree of 8 leaves. Left Panel: labels for Modified Laplacian version. The cherry formed by nodes (13,14,15) is replaced by a single leaf labeled as 13. Right Panel: labels for Modified Reduced Laplacian matrices. The leaf 8 was removed. Black ticks divide branches into parts of unit length.}
 \label{trees_maxbal71}
\end{figure*}

\ \ \ \ \ There is at least one maximally balanced tree for any value of $n$. Every fully balanced tree is also maximally balanced, but the opposite is not true. To extend the results of previous section to trees of arbitrary sizes, we follow the algorithm presented in \citep{fis:2019} and build the Distance matrix $\boldsymbol{W}^{(s,n)}$ and $\boldsymbol{W}^{(s,n)}_e$ for values of $n \in (2^{s-1},2^s)$. The construction starts with a fully balanced tree with size $n'=2^s$ and finds a cherry of maximal depth in the tree, that is, the maximal value of the first line of $\boldsymbol{W}^{(s)}$  (or any cherry in the reduced version). Then, the cherry tips are removed from the tree so that the corresponding internal node becomes a tip with a new branch of length two from its ancestor. These two steps generate a tree of size $n'=2^s-1$. This process is repeated until $n'$ reaches the desired value of $n$. Figures \ref{trees_maxbal71} show examples of trees and their respective distance matrices. In Appendix \ref{AA7} we show examples for $s=3$ and $n=6$ and 5. The Modified Degrees Matrix $\boldsymbol{D}^{(s,n)}$ and its reduced version $\boldsymbol{D}^{(s,n)}_e$ can be obtained from Eq.(\ref{degree}) and the Normalized Modified Laplacian matrix can be written down.

\begin{figure*}[!htpb]
 \centering
 \includegraphics[width=0.75\columnwidth]{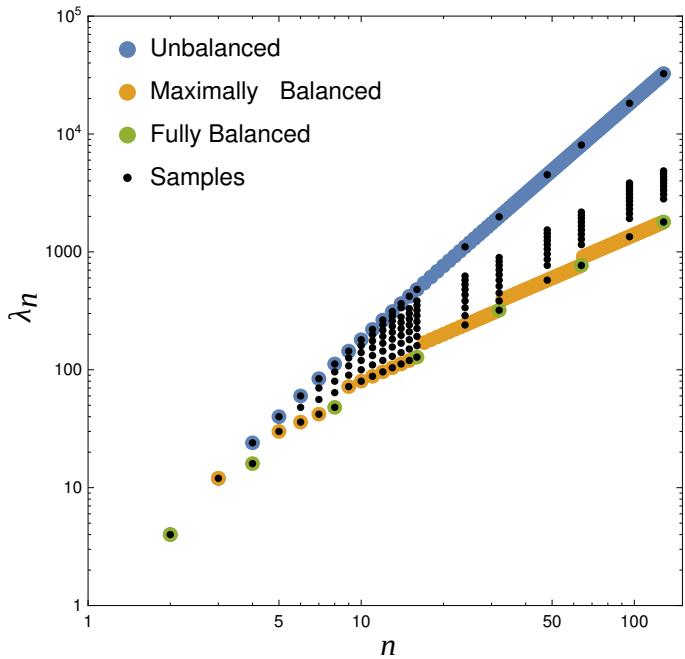}
 \caption{The behavior of the largest eigenvalue as a function of $n$. For large $n$, $\lambda_n \simeq n^2$ in the fully unbalanced case and $\lambda_n \simeq n\log{n}$ in the balanced cases. The eigenvalues associated with unbalanced trees are always larger than the balanced ones for $n \geq 4$. The small dots are the largest eigenvalues for samples of trees constructed randomly.}
 \label{lambplot}
\end{figure*}

\subsubsection{Spectrum}

\ \ \ \ \ We do not have general expressions for all eigenvalues of $\boldsymbol{L}_e$ in this case. However, the largest eigenvalue is given by (see Appendix \ref{AA6}):
\begin{equation}
 \label{lamb_maxbalf2}
 \lambda_{n}=2sn, \qquad n \in (2^{s-1},2^s).
\end{equation}
Figure \ref{lambplot} shows these largest eigenvalues together with the fully balanced and unbalanced cases. The most remarkable difference between the eigenvalues of fully unbalanced and the balanced cases is the presence of the multiplicities. Except for the zero and the largest eigenvalue, all eigenvalues of fully balanced trees are degenerated with different multiplicities, whereas in the unbalanced case, there are no degeneracies. Another important point concerns the largest eigenvalue $\lambda_n$, which is always larger for unbalanced trees for $n \geq 4$. The largest eigenvalue of maximally balanced trees jumps up from a fully balanced case and tends smoothly to the next fully balanced tree.\\

The calculation of the largest eigenvalue $\lambda_n$ for these extreme cases, does not prove that their values are extreme (i.e., there might be trees of the same size with $\lambda_n$ larger than the fully unbalanced case or $\lambda_n$ smaller than the balanced case.) To investigate this property and establish some results in this direction, we constructed samples of trees following the Yule random model \citep{har:1971,kir:1993,ste:2001,car:2013} and calculated the largest eigenvalue of each tree in these sets. For $n \in [2,8]$ the number of trees is $(n-1)!$. For $n \geq 9$, the total number of trees is of the order of $(4 \cdot 10^4 / n)$. In our numerical calculations, we never found cases where the largest eigenvalue was larger than the fully unbalanced case or smaller than maximally balanced cases. This indicates that the Laplacian spectra of fully balanced and maximally balanced cases are extremal and can serve as scales to measure balance of phylogenetic trees. However, different trees can have the same $\lambda_n$ and there are distributions of the largest eigenvalue for each $n$. This is also true of  the Sackin index \citep{fis:2019}.

\section{Conclusions}
\label{conc}

\ \ \ \ \ Powerful statistical results provide evidence of the monophyly of all known life on Earth \citep{ste:2010,the:2010}, supporting the hypothesis of a Universal Common Ancestry (UCA). Even if life has started multiple times independently, the descendants of only one of the attempts have survived, exclusively, until the present. According to this monophyletic hypothesis, all living species have a single common ancestor that lived over a billion years ago. Understanding how this phylogenetic tree evolved and how its structural properties relate to evolutionary processes is an open question.\\

In this paper, we constructed Laplacian matrices for extreme forms of phylogenetic trees, namely, fully unbalanced, fully balanced, and maximally balanced. By inspecting a few particular cases, we were able to find all the matrix elements of the Distance matrix $\boldsymbol{W}$ for any number $n$ of extant species in the unbalanced case. For fully balanced and maximally balanced trees, we introduced recursive rules that also allowed us to construct  $\boldsymbol{W}$ for $n=2^s$ and integer $s$ and $\boldsymbol{W}$ for $n \in (2^{s-1},2^s)$ respectively. From these results, we calculated the Modified Normalized Laplacian matrix $\boldsymbol{L}=\boldsymbol{D}-\boldsymbol{W}$ and its normalized version $\boldsymbol{L}_n=\boldsymbol{I}-\boldsymbol{W}_n$. The same procedures were applied to construct a reduced version of these matrices, where only extant species are used as nodes. In the entire calculation, branch lengths have been normalized as multiples of one, to focus on the tree balance. Changing branch lengths would probably change the eigenvalues and might turn exact degeneracies (in the case of fully balanced trees) into clusters of nearby eigenvalues. We have not, however, analyzed this situation, as it would break the perfect pattern of the Laplacian matrix and would probably be very hard to study analytically.\\

We found that the Distance matrix is self-similar for the fully balanced trees, which is a consequence of the trees' self-similarity. Self-similar patterns emerge when a population breaks in two species, and these two species break again in two, and so on. In neutral models of speciation this type of process is more likely to happen in sympatry, when space is not relevant \citep{agu:2017}. In spatial populations, simulations show that species typically emerge when a spatially localized group of individuals branches off the original population, followed by a second group and so on, giving rise to a caterpillar type tree \citep{cos:2019}. The spectrum of the Laplacian matrix might help understand how these different processes are reflected in the phylogenetic trees.\\

We computed all eigenvalues of the reduced Laplacian matrix for  fully unbalanced and balanced cases, equations (\ref{lambr_unbf2}) and (\ref{lambs_balf2}) respectively. We find a remarkable difference between the spectra, namely, the presence of multiplicities when the tree is balanced. Fully unbalanced trees have no degeneracy, and each eigenvalue is unique. The spectral density of balanced trees must, therefore, exhibit peaks at highly degenerated eigenvalues  whereas unbalanced trees should display much smoother densities. Although exact degeneracies are likely broken by deviations from the fully balanced structure, peaks should remain if the eigenvalues remain sufficiently close together.\\

For maximally balanced trees we did not find the complete set of eigenvalues and focused on the largest eigenvalue of $\boldsymbol{L}_e$ only. This was calculated for all $n \in (2^{s-1},2^s)$.  The largest eigenvalue of a fully unbalanced tree is always larger than the largest eigenvalue of a fully balanced tree of equal size. We tested this extremal property of fully unbalanced and maximally balanced trees computing the largest eigenvalue for trees constructed randomly with the Yule model,  Figure \ref{lambplot}. We believe these results will contribute to recent analysis of phylogenies from the spectral density profile of Laplacian matrices \citep{lew:2016,lew:2016b,lew:2020,mor:2016,lew:2019,chi:2019,mor:2019,dea:2020,yer:2020}.

\bibliographystyle{spbasic}      
\bibliography{refs_2020_BMAB_AraujoLimaAguiar_LaplacianMatrices.bib}   

\appendix
\appendixpage
\section{Fully Unbalanced Tress - Modified Matrix Version}
\label{AA1}

\ \ \ \ \ We will construct expressions for $w_{i,j}^{(n)}$ by inspecting the matrices in Eq.(\ref{Ls_unb}). For a cherry, consisting in two extant species and one internal node, we find:
\begin{equation}
 \label{Ls_unb_A}
 \boldsymbol{L}^{(2)} =\begin{pmatrix}
  2 & -1 & -1\\
  -1 & 3 & -2\\
  -1 & -2 & 3\end{pmatrix}.
\end{equation}
By simplicity, we start with the odd nodes. Since these are internal nodes, the distance between odd nodes depends only on their labels. Generally, if $i$ and $j$ are odd, the matrix element is
\begin{equation}
 \label{wij_odd_unbal}
 w_{i,j}^{(n)}=\frac{j-i}{2}.
\end{equation}
The distance is independent of $n$, since odd nodes do not change their location as $n$ increases. If $i$ is odd, but $j$ is even the distance between them does increase with $n$. However, $w_{i,j}^{(n)}$ does not depend on $j$, recalling that $j>i$. When $i$ increases, the distance to even $j$ decreases by 1, and we can write
\begin{equation}
 w_{i,j}^{(n)} = n-1-\frac{i-1}{2} = n-\frac{i+1}{2}.
  \label{wij_odd_even_unbal}
\end{equation}
If both nodes are even the distance increases with $2n$, since the two nodes depart from the root simultaneously. Fixing a even node $i$, the distance $w_{i,j}^{(n)}$ is, for $j>i$:
\begin{equation}
 \label{wij_even_unbal}
 w_{i,j}^{(n)} = 2n-i.
\end{equation}
Finally, we must obtain $w_{i,j}^{(n)}$ for $i$ even and $j$ odd. Fixing $i$ the distance increases with $n$ and $j$, as the next odd node is more distant from $i$ than the previous one. If $i$ increases, on the other hand, the distance decreases by 1. Inspecting the matrices on (\ref{Ls_unb}) and (\ref{Ls_unb_A}) we find
\begin{equation}
 \label{wij_even_odd_unbal}
 w_{i,j}^{(n)}=n+\frac{j-1}{2}-(i-1) = n-i+\frac{j+1}{2}.
\end{equation}
Because parity plays a key role, we first calculate the matrix elements for odd $i$:
\begin{equation}
 \label{di_odd0}
 d_i^{(n)}=\sum_{j=1}^{n} w_{i,2j-1}^{(n)}+\sum_{j=1}^{n-1} w_{i,2j}^{(n)}
\end{equation}
where we split the sum into two parts, for odd and even $j$, respectively. Because $w_{i,j}^{(n)}$ depends on whether $i<j$ or $i>j$ we split the sums again as
\begin{eqnarray}
 \label{di_odd1}
 d_i^{(n)}=\sum_{j=1}^{(i-1)/2} w_{i,2j-1}^{(n)} + \sum_{j=(i+3)/2}^{n} w_{i,2j-1}^{(n)} + \nonumber\\
 \sum_{j=1}^{(i-1)/2} w_{i,2j}^{(n)} + \sum_{j=(i+1)/2}^{n-1} w_{i,2j}^{(n)},
\end{eqnarray}
or
\begin{eqnarray}
 \label{di_odd2}
 2d_i^{(n)}=\sum_{j=1}^{(i-1)/2} (i-2j+1) + \sum_{j=(i+3)/2}^{n} (2j-i-1) + \nonumber\\
 \sum_{j=1}^{(i-1)/2} (2n-4j+i+1) + \sum_{j=(i+1)/2}^{n-1} (2n-i-1).
\end{eqnarray}
All sums are arithmetic and can be calculated. A similar procedure works for even $i$ resulting in
\begin{equation}
 \label{di_unbalf_A}
 2d_i^{(n)} = \left\{ \begin{array}{l}
  3n^2-n(2i+3)+i(i+1), \quad i \text{ odd}\\
  7n^2-n(4i+7)+i(i+2), \quad i \text{ even.}\end{array}\right.
\end{equation}

\section{The Laplacian Matrix is Positive Semi-definite}
\label{AA2}

\ \ \ \ \ We will show that $\boldsymbol{L}_e$ is a positive semi-definite matrix, i.e., $\boldsymbol{x}^t \boldsymbol{L}_e \boldsymbol{x} \geq 0$, for any vector $\boldsymbol{x}$ and its transpose $\boldsymbol{x}^t$. If the matrix elements of $\boldsymbol{L}_e$ are
\begin{equation}
 \label{lij}
 \ell_{i,j} = \left\{ \begin{array}{l}
  \sum_{\substack{q=1 \\ q \neq i}}^n w_{i,q}, \quad i=j\\
  -w_{i,j}, \quad i \neq j,\end{array}\right.
\end{equation}
and $w_{i,j}$ are the elements of the Distance matrix:
\begin{equation}
 \label{vLv}
 \boldsymbol{x}^t \boldsymbol{L}_e \boldsymbol{x} = \sum_{p=1}^n \sum_{q=1}^n x_p x_q \ell_{p,q}
\end{equation}
\begin{equation*}
 = \sum_{p=1}^n x_p^2 \ell_{p,p} + \sum_{p=1}^n \sum_{\substack{q=1 \\ q \neq p}}^n x_p x_q \ell_{p,q}
\end{equation*}
\begin{equation*}
 = \sum_{p=1}^n x_p^2 \sum_{\substack{q=1 \\ q \neq p}}^n w_{p,q} - \sum_{p=1}^n \sum_{\substack{q=1 \\ q \neq p}}^n x_p x_q w_{p,q}
\end{equation*}
\begin{equation}
 \label{vLvw}
 \boldsymbol{x}^t \boldsymbol{L}_e \boldsymbol{x} = \sum_{p=1}^n \sum_{\substack{q=1 \\ q \neq p}}^n x_p (x_p - x_q) w_{p,q}.
\end{equation}
These last sums can be split into terms with $p > q$ and $p < q$:
\begin{equation*}
  \boldsymbol{x}^t \boldsymbol{L}_e \boldsymbol{x} = \sum_{p>q} x_p (x_p - x_q) w_{p,q} + \sum_{p<q} x_p (x_p - x_q) w_{p,q}
\end{equation*}
\begin{equation}
 \label{vLvqp}
 \boldsymbol{x}^t \boldsymbol{L}_e \boldsymbol{x} = \sum_{p>q} x_p (x_p - x_q) w_{p,q} + \sum_{q<p} x_q (x_q - x_p) w_{q,p}
\end{equation}
How the Distance matrix is symmetric, $w_{p,q} = w_{q,p} > 0$:
\begin{equation}
 \label{vLvfinal}
 \boldsymbol{x}^t \boldsymbol{L}_e \boldsymbol{x} = \sum_{p>q} (x_p - x_q)^2 w_{p,q} \geq 0.
\end{equation}
Demonstrating that $\boldsymbol{L}_e$ is a semi-positive definite matrix.

\section{Laplacian Matrix Spectrum - Fully Unbalanced Case}
\label{AA3}

\ \ \ \ \ For the fully unbalanced case, the sets of eigenvalues corresponding some $\boldsymbol{L}_e^{(n)}$ are 
\begin{equation}
 \label{lamb_unb}
 \left\{ \begin{array}{l}
  n=2 \Rightarrow \{\lambda_i\}_{i=1}^2 = \{0,4\},\\
  n=3 \Rightarrow \{\lambda_i\}_{i=1}^3 = \{0,8,12\},\\
  n=4 \Rightarrow \{\lambda_i\}_{i=1}^4 = \{0,14,18,24\},\\
  n=5 \Rightarrow \{\lambda_i\}_{i=1}^5 = \{0,22,26,32,40\}.\end{array}\right.
\end{equation}
By inspection, we see that the largest eigenvalue is $\lambda_n=2d_n^{(n)}$ and from (\ref{di_unbal_l2}) we get
\begin{equation}
 \label{lambn_unb}
 \lambda_n = 2n(n-1).
\end{equation}
The remaining eigenvalues can be expressed recursively from the largest one:
\begin{equation}
 \label{lambn_unb_rec}
 \lambda_{n-1} = \lambda_n-2(n-1),
\end{equation}
\begin{equation*}
 \lambda_{n-2} = \lambda_{n-1}-2(n-2),
\end{equation*}
\begin{equation*}
 \lambda_{n-3} = \lambda_{n-2}-2(n-3),
\end{equation*}
\begin{equation*}
 \vdots
\end{equation*}
This can also be written as 
\begin{equation}
 \label{lambr_unb}
 \lambda_{n-r} = \lambda_{n-r+1}-2(n-r), \qquad r=1,2,3,...,n-2,
\end{equation}
\begin{equation*}
 \lambda_{n-2} = \lambda_n-2(n-1)-2(n-2),
\end{equation*}
\begin{equation*}
 \lambda_{n-3} = \lambda_n-2(n-1)-2(n-2)-2(n-3),
\end{equation*}
\begin{equation*}
 \lambda_{n-r} = \lambda_n-2\sum_{q=1}^r (n-q)=\lambda_n-r(2n-1-r),
\end{equation*}
\begin{equation}
 \lambda_{n-r} =\lambda_n-r(2n-1-r), \qquad r=1,2,3,...,n-2.
\end{equation}
Making $n-r \rightarrow r+1$ and replacing $\lambda_n$ into last expression, we get
\begin{equation}
 \label{lambr_unbf}
 \lambda_{r+1} = n(n-1)+r(r+1), \qquad r=1,2,3,...,n-1.
\end{equation}

\begin{figure*}[!htpb]
 \centering
 \includegraphics[width=0.9\textwidth]{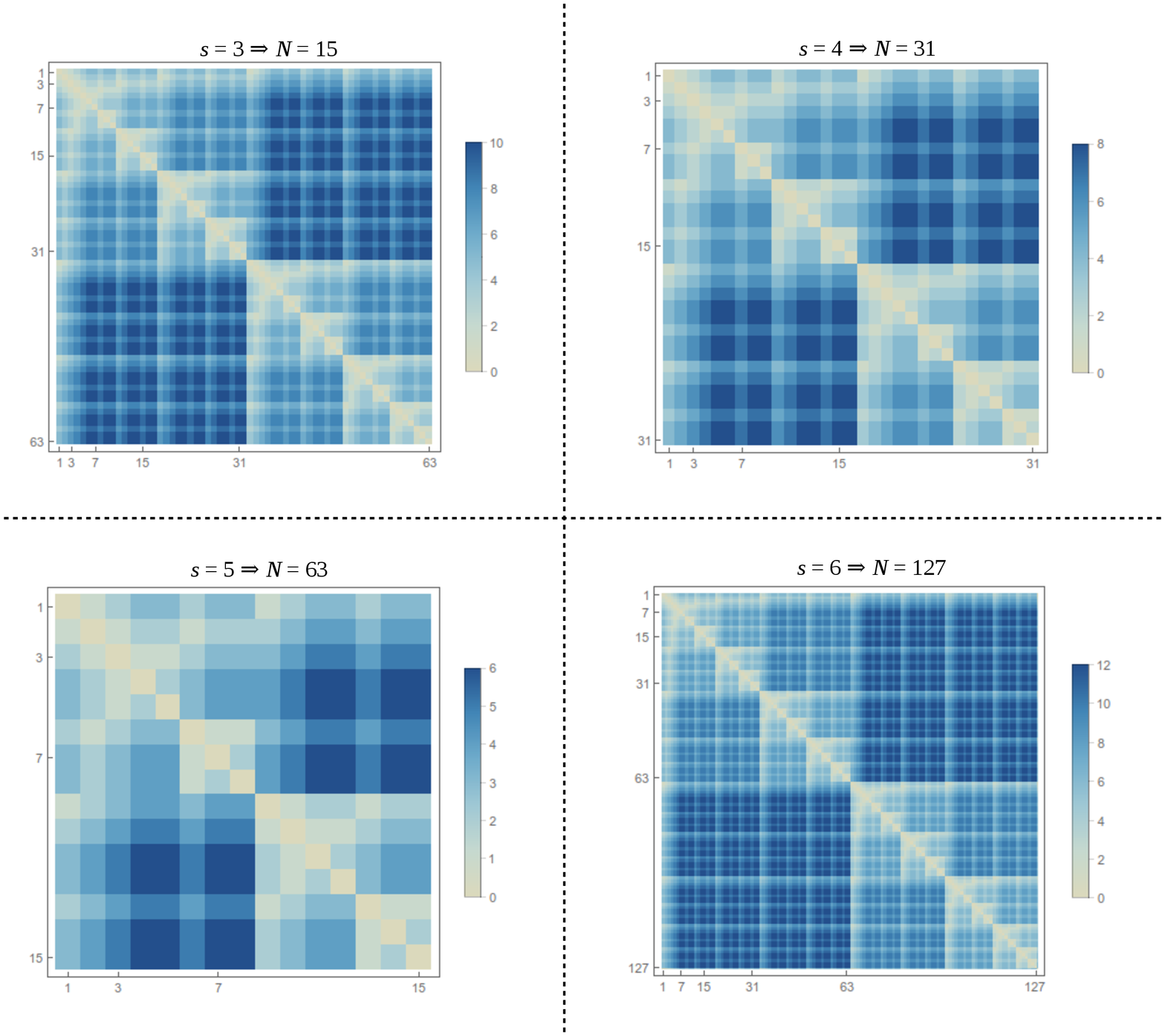}
 \caption{Self-similar pattern of the distance matrix $\boldsymbol{W}^{(s)}$ for the fully balanced trees and some depths. Dark blue cells indicate large distances while the light blue ones indicate small distances. The large distances are concentrated in the filling sub-matrices $\boldsymbol{W}^{(s)}_F$, distances between nodes on opposite sides of the root node. The small distances are concentrated in regions that come from distance matrix $\boldsymbol{W}^{(s-1)}$, distances between nodes on the same side of the root node.}
 \label{mat_plot_A}
\end{figure*}

\begin{figure*}[!htpb]
 \centering
 \includegraphics[width=0.9\textwidth]{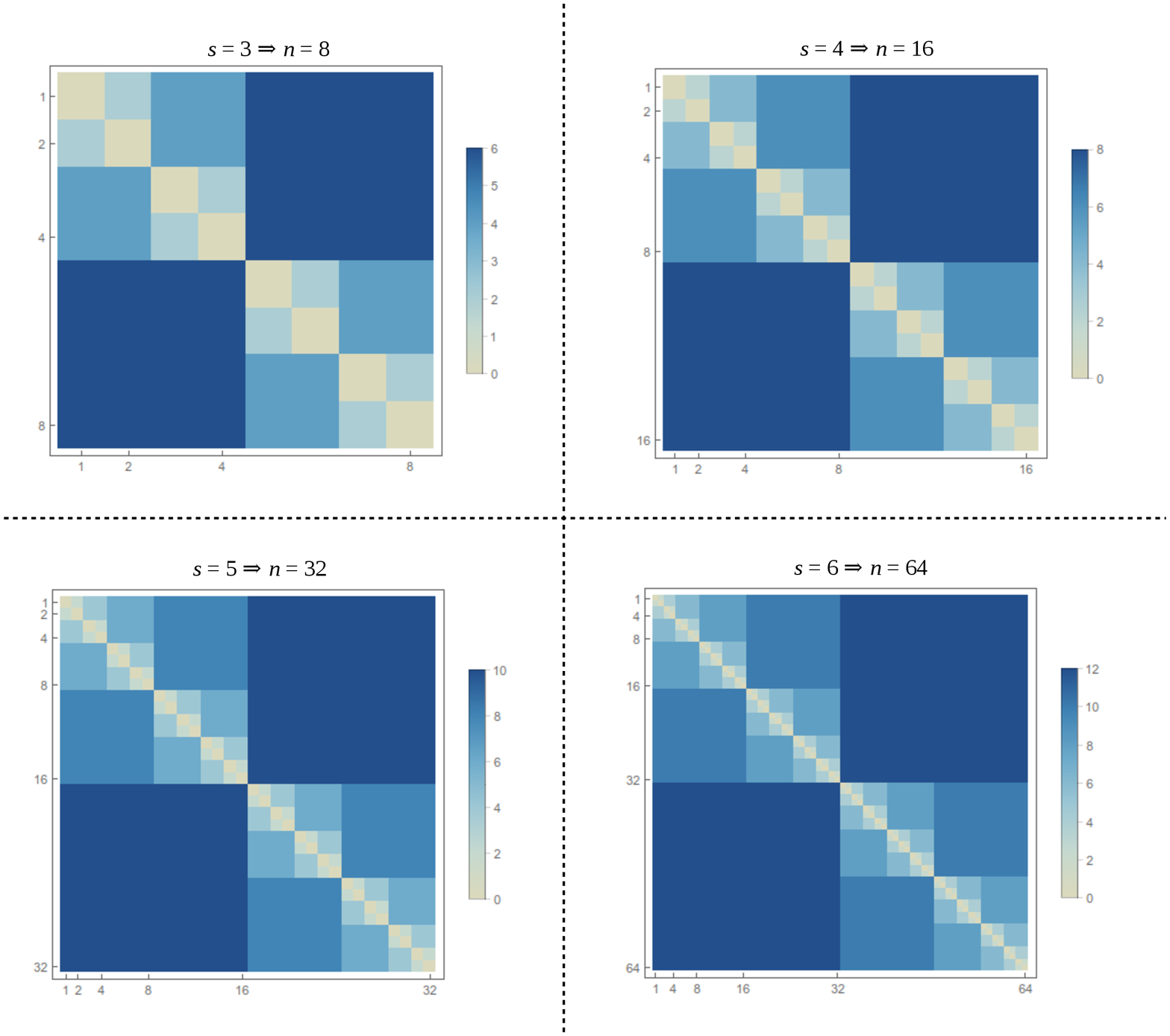}
 \caption{Self-similar pattern of the distance matrix $\boldsymbol{W}^{(s)}_e$ for the fully balanced trees and some depths. Dark blue cells indicate large distances while the light blue ones indicate small distances. The large distances are concentrated in the filling sub-matrices $\boldsymbol{W}^{(s)}_{e,F}$, distances between nodes on opposite sides of the root node. The small distances are concentrated in regions that come from distance matrix $\boldsymbol{W}^{(s-1)}_e$, distances between nodes on the same side of the root node.}
 \label{mat_plot_A2}
\end{figure*}

\section{Fully Balanced Trees - Modified Matrix Version}
\label{AA4}

\ \ \ \ \ Let us look at the first line. To shed light on analysis, we show below the first line of for $s=4$:
\begin{eqnarray}
 \label{W1}
 (\boldsymbol{W}^{(4)})_{i=1} = (0 \quad 1 \quad 2 \quad 3 \quad 4 \quad 4 \quad 3 \quad 4 \quad 4 \quad 2 \quad 3 \quad 4 \quad 4 \nonumber\\
 \textcolor{red}{1} \quad 2 \quad 3\quad 4 \quad 4 \quad 3 \quad 4 \quad 4 \quad 2 \quad 3 \quad 4 \quad 4).
\end{eqnarray}
The matrices in Eqs. (\ref{Ls_unb_A}), (\ref{Ls_bal}), (\ref{W1}) and Figure \ref{mat_scheme} have 4 different sizes. In all of them there is a counting from 1 to $s$, then $s$ repeats. From this repetition starts a new counting from $s-1$ to $s$, then $s$ repeats. From this a new series starts counting from $s-2$ to $s$, then $s$ repeats. This goes on until the counting starts again at 1, red highlight in (\ref{W1}), and the line is filled by the same previous elements. We must follow a set of rules that describes the counting in the first line. For example, for $s=4 \Rightarrow n=16$ the rules are
\begin{equation}
 \label{s4r}
 w_{1,j}^{(4)}=\left\{ \begin{array}{l}
  j-1, \quad 1 \leq j \leq 5\\
  4, \quad\qquad j=6\\
  w_{1,j-3}^{(4)}, \quad 7\leq j\leq 9\\
  w_{1,j-7}^{(4)}, \quad 10\leq j\leq 16\\
  w_{1,j-15}^{(4)}, \quad 17\leq j\leq 31.\end{array}\right.
\end{equation}
In general, we have the first 2 rules plus $s-1$ additional ones. To write them we need intervals $j_{\text{min}} \leq j \leq j_{\text{max}}$ and a value due to restart the counting. By inspection we arrive at the following set of rules for $r=1,2,...,s-1$:
\begin{eqnarray}
 \label{sr}
 w_{1,j}^{(s)}=\left\{ \begin{array}{l}
  j-1, \quad 1 \leq j\leq s+1\\
  s, \quad\qquad j=s+2\\
  w_{1,j+1-2^{r+1}}^{(s)}, \quad j_{\text{min}}(r) \leq j \leq j_{\text{max}}(r)\end{array}\right.
\end{eqnarray}
where the range delimiters $j_{\text{min}}(r)$ and $j_{\text{max}}(r)$ are given by
\begin{equation}
 \label{jrs}
 \left \{ \begin{array}{l}
  j_{\text{min}}(r) = s+3+\sum_{q=2}^r (2^q-1)\\
  j_{\text{max}}(r) = s+5+\sum_{q=2}^r (2^{q+1}-1),\end{array}\right.
\end{equation}
and result in
\begin{equation}
 \label{jr}
 \left \{ \begin{array}{l}
  j_{\text{min}}(r) = s-r+2^{r+1}\\
  j_{\text{max}}(r) = j_{\text{min}}(r) + 2(2^r-1).\end{array}\right.
\end{equation}
We now consider the filling sub-matrices $\boldsymbol{W}_F^{(s)}$. The information contained in these matrices is the distances between the first half of nodes (except the root node 1) and the second half of nodes. Each element is a measure of the distance between nodes on opposite sides of the root node 1. So the distance between these nodes depends on the distance between these nodes and the root. For $j>i$ and $w_{i,j}^{(s)} \in \boldsymbol{W}_F^{(s)}$, i.e. $2\leq i \leq 2^s$ and $2^s+1\leq j \leq 2^{s+1}-1$, we have
\begin{equation}
 \label{wf}
 w_{i,j}^{(s)}=w_{1,i}^{(s)} + w_{1,j}^{(s)}.
\end{equation}
Thus the Distance matrix $\boldsymbol{W}^{(s)}$ is completely defined for any value of $s$. The previously mentioned self-similar pattern is presented in Figures \ref{mat_plot}, \ref{mat_plot_A}, and \ref{mat_plot_A2}.\\

To calculate the modified degrees of internal nodes and tips, we recall that all $\boldsymbol{L}^{(s)}$ matrices is divided into two identical blocks, for example, the red sub-matrices in Figure \ref{mat_scheme}. Therefore, we only need to calculate the modified degrees of half the nodes,  i.e., for $i \leq 2^s$:
\begin{eqnarray}
 \label{di_bal}
 d_i^{(s)} = w_{i,1}^{(s)} + \sum_{j=2}^{i-1} w_{i,j}^{(s)} + \sum_{j=i+1}^{2n-1} w_{i,j}^{(s)} = \nonumber\\
 w_{i,1}^{(s)} + \sum_{j=2}^{i-1} w_{i,j}^{(s)} + \sum_{j=i+1}^{2^s} w_{i,j}^{(s)} + \sum_{j=2^s+1}^{2^{s+1}-1} w_{i,j}^{(s)}.
\end{eqnarray}
The initial sum in (\ref{degree}) was split into four parts. The first one is the distance $w_{i,1}^{(s)}$. The following two are the sum of terms in $\boldsymbol{W}^{(s-1)}$ and the last one is the sum of terms in $\boldsymbol{W}_F^{(s)}$. This last term is easy to sum
\begin{equation*}
 \sum_{j=2^s+1}^{2^{s+1}-1} w_{i,j}^{(s)} = \sum_{j=2^s+1}^{2^{s+1}-1} (w_{1,i}^{(s)} + w_{1,j}^{(s)}),
\end{equation*}
\begin{equation}
 \label{di_bal_wf}
 \sum_{j=2^s+1}^{2^{s+1}-1} w_{i,j}^{(s)} = [w_{1,i}^{(s)}(2^s-1)] + [1+2^s(s-1)].
\end{equation}
The two terms coming from $\boldsymbol{W}^{(s-1)}$ in (\ref{di_bal}) are exactly the modified degree measured for $s-1$ for the node $i-1$, because for the next $s$ the matrix is constructed from the second line. Assembling all these terms and arranging we obtain
\begin{equation}
 \label{di_bal_final}
 d_i^{(s)} = \left\{ \begin{array}{l}
  d_{i-1}^{(s-1)} + 2^s(w_{1,i}^{(s)}+s-1) + 1, \quad 2 \leq i \leq 2^s\\
  d_{i+1-2^s}^{(s)}, \quad 2^s+1 \leq i \leq 2^{s+1}-1.\end{array}\right.
\end{equation}

\section{Laplacian Matrix Spectrum - Fully Balanced Case}
\label{AA5}

\ \ \ \ \ For the fully balanced case, the sets of eigenvalues corresponding some $\boldsymbol{L}_e^{(s)}$ are
\begin{equation}
 \label{lamb_bal}
 \left\{ \begin{array}{l}
  s=1 \Rightarrow \{\lambda_i\}_{i=1}^2 = \{0,4\},\\
  s=2 \Rightarrow \{\lambda_i\}_{i=1}^4 = \{0,12(2),16\},\\
  s=3 \Rightarrow \{\lambda_i\}_{i=1}^8 = \{0,36(4),40(2),48\},\\
  s=4 \Rightarrow \{\lambda_i\}_{i=1}^{16} = \{0,100(8),104(4),112(2),128\}.\end{array}\right.
\end{equation}
where the numbers in parenthesis are the multiplicities of each eigenvalue (unit degeneracies are not labeled). In all examples, the first non-zero eigenvalue is always $d^{(s)}+2$ and its multiplicity is $2^{s-1}$. The next different eigenvalue is the previous one plus $4$ and the next one plus $8$:
\begin{eqnarray}
 \label{lambs_bal}
 \{\lambda_i\}_{i=1}^{2^s}=\{0,d^{(s)}+2(2^{s-1}),d^{(s)}+2+4(2^{s-2}),..., \nonumber\\
 d^{(s)}+2+4+...+2^s\}.
\end{eqnarray}
Each eigenvalue is
\begin{equation*}
 \lambda_r = d^{(s)} + 2^1 + 2^2 + 2^r = d^{(s)} + \sum_{m=1}^r 2^m = d^{(s)} + 2(2^r-1),
\end{equation*}
\begin{equation}
 \label{lambs_bal2}
 \lambda_r = 2[1+2^s(s-1)] + 2(2^r-1),
\end{equation}
and appears with multiplicity $2^{n-r}$. Thus, the set of eigenvalues of $\boldsymbol{L}_e^{(s)}$ contains 0, which appears only once, and
\begin{equation}
 \label{lambs_balf}
 \lambda_r = 2 [2^r+2^s(s-1)], \quad r=1,2,3,...,s,
\end{equation}
with multiplicity $2^{s-r}$. Where the largest eigenvalue in this case is
\begin{equation}
 \label{lambs_balmax}
 \lambda_s = s2^{s+1}.
\end{equation}

\section{Laplacian Matrix Largest Eigenvalue - Maximally Balanced Case}
\label{AA6}

\ \ \ \ \ For the maximally balanced case, the sets of eigenvalues corresponding to same $\boldsymbol{L}_e^{(s,n)}$ are
\begin{equation}
 \label{lamb_maxbal}
 \left\{ \begin{array}{l}
  n=5 \Rightarrow \{\lambda_i\}_{i=1}^5 = \{0,20,24,26,30\},\\
  n=6 \Rightarrow \{\lambda_i\}_{i=1}^6 = \{0,24(2),28,32,36\},\\
  n=7 \Rightarrow \{\lambda_i\}_{i=1}^7 = \{0,30(2),32,34,36,42\}.\end{array}\right.
\end{equation}
We do not have general expressions for all eigenvalues in this case. However, the largest eigenvalue obeys a general expression that we will also show by inspection. The largest eigenvalues for $n \in (8,16)$ are:
\begin{equation}
 \label{lamb_maxbalmax}
 \{\lambda_n\}_{n=9}^{15} = \{72,80,88,96,104,112,120\}.
\end{equation}
By inspecting the three intervals, $n \in (2,4)$, (4,8) and (8,16), we observe that the largest eigenvalue is exactly
\begin{equation*}
 \lambda_{15}=128-8,
\end{equation*}
\begin{equation*}
 \lambda_{15}=128-2 \cdot 4,
\end{equation*}
\begin{equation*}
 \lambda_{2^s-1}=\lambda_{2^s}-2s,
\end{equation*}
\begin{equation*}
 \lambda_{2^s-2}=\lambda_{2^s}-4s,
\end{equation*}
\begin{equation*}
 \vdots
\end{equation*}
\begin{equation*}
 \lambda_{2^s-p}=\lambda_{2^s}-2ps,
\end{equation*}
\begin{equation}
 \label{lamb_maxbal2}
 \lambda_{2^s-p}=2s(2^s-p), \qquad p=1,2,...,2^{s-1}-1,
\end{equation}
\begin{equation}
 \label{lamb_maxbalf}
 \lambda_{n}=2sn, \qquad n \in (2^{s-1},2^s).
\end{equation}

\newpage

\section{Balanced Trees - Distances Matrices}
\label{AA7}

\begin{figure*}[!htpb]
 \centering
 \includegraphics[width=0.9\columnwidth]{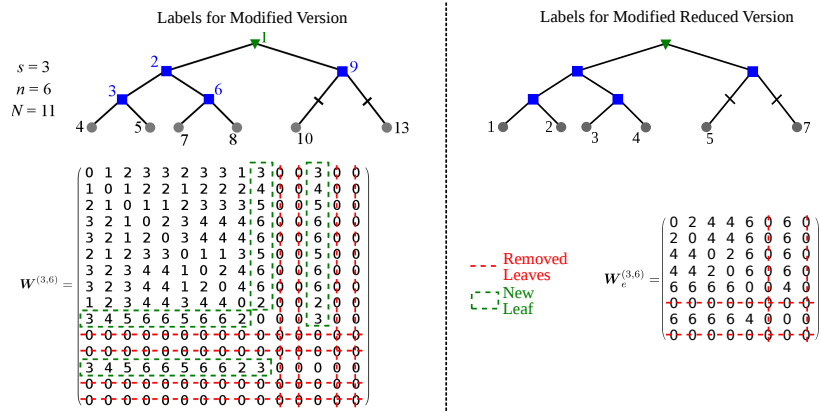}
 \caption{Maximally balanced tree with six tips. Two steps reach the desired tree starting the algorithm with a tree of 8 leaves. Left Panel: labels for Modified Laplacian version. The cherries formed by nodes (13,14,15) and (10,11,12) are replaced by leaves labeled as 13 and 10. Right Panel: labels for Modified Reduced Laplacian matrices. The leaves 8 and 6 were removed. Black ticks divide branches into parts of unit length.}
 \label{trees_maxbal6}
\end{figure*}
\begin{figure*}[!htpb]
 \centering
 \includegraphics[width=0.9\columnwidth]{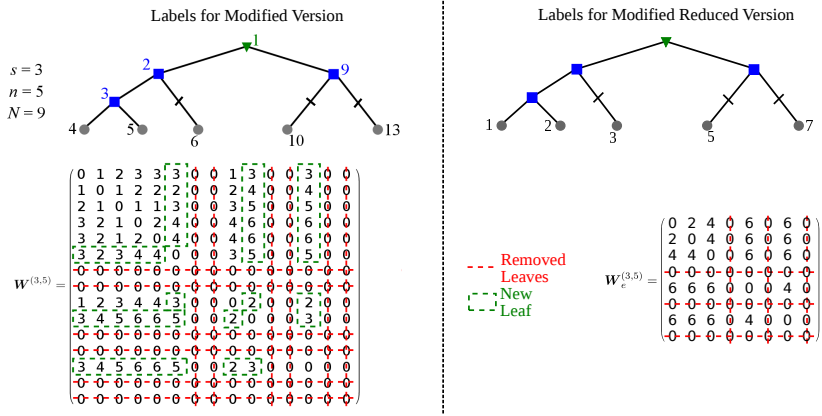}
 \caption{Maximally balanced tree with five tips. Three steps reach the desired tree starting the algorithm with a tree of 8 leaves. Left Panel: labels for Modified Laplacian version. The cherry formed by nodes (13,14,15), (10,11,12) and (6,7,8) are replaced by leaves labeled as 13, 10 and 6. Right Panel: labels for Modified Reduced Laplacian matrices. The leaves 8, 6 and 4 were removed. Black ticks divide branches into parts of unit length.}
 \label{trees_maxbal5}
\end{figure*}

\end{document}